# Big Data : Overview

Richa Gupta[1], Sunny Gupta[2], Anuradha Singhal[3]
[13](Department of Computer Science, University of Delhi, India)
[2](University of Delhi, India)

***ABSTRACT:*** *Big data is data that exceeds the processing capacity of traditional databases. The data is too big to be processed by a single machine. New and innovative methods are required to process and store such large volumes of data. This paper provides an overview on big data, its importance in our live and some technologies to handle big data.*

## I.  INTRODUCTION

There's a lot of hype around today about big data. The question comes in what is big data and why is it being considered so important. Before answering this question, we should first know how our so-called traditional approaches have been. The traditional methods of analysis requires two things: 1) that data to be of structured form, that is, one which can fit easily in relational database. 2) the analysis to be done by a software residing on a single machine.

But, as the trends are changing so is the nature of data that is been produced. Today, most common form of data is images and text. This type of data finds it difficult to fit in relational database which makes it difficult to analyse using traditional approach. Big data in turn is required to handle structured, semi-structured and unstructured data. Structured data can fit easily in a relational database and can be processed using simple to complex SQL queries. Semi-structured data is one that doesn't fit into database but have some organizational properties that make it easy to analyse. Eg: XML and NoSQL databases. Unstructured data is like videos, images, text, presentations, audio files, web pages. They do not fit neatly into the database.

Not only does big data involves structured and unstructured data, it is also "huge".  The data is in not in terms of megabytes or terabytes but as large as petabytes and zetabytes, which is further going to increase. Big data is data which is huge for any company, distributed on several machines. Its size makes it natural in the involvement of many complex data structures to analyse the big data.

## II.   IMPORTANCE

The ultimate goal of big data is to provide as output some business solutions that can help company to gain in business solutions.  This reason itself makes the analysis of big data important. For example any company can benefit if they could figure out that if customer buys "X" then it is likely that he/she would be interested in buying "Y" also. This type of analysis at run-time can greatly benefit by increasing business. Social networking sites are analysing web logs to recommend users what he/she might be interested in. Big data aims at dramatic cost reduction and substantial improvements in the time required to perform a computing task.

## III.   III. KEY ISSUES

There are three main keys for big data

Volume -  the size of data now is larger than terabytes and petabytes. The grand scale and rise of size makes it difficult to store and analyse using traditional tools.  For example,  Facebook ingests 500 terabytes of data every day.

Velocity – for time limited processes, big data should be used as it streams data in order to maximize its value.

Variety – Big data comes from a variety of sources. Traditional database systems were designed to address smaller volumes of structured data, fewer updates or a predictable, consistent data structure whereas Big Data is also geospatial data, 3D data, audio and video, and unstructured text, including log files and social media[1].

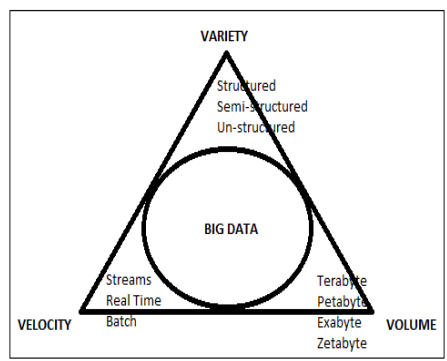

**Figure 1: Keys of Big Data**

## IV. BIG DATA SAMPLES

Big data samples are available in astronomy, atmospheric science, social networking sites, life sciences, medical science, government data, natural disaster and resource management, web logs, mobile phones, sensor networks, scientific research, telecommunications.

## V. TECHNOLOGIES FOR BIG DATA

Big data has great potential to produce useful information for companies which can benefit the way they manage their business solutions. MapReduce is a programming model for processing lagre data sets with a parallel, distributed algorithm on a cluster [2]. MapReduce is divided into two broad steps:

- Map : Mapper performs the task of filtering and sorting
- Reduce : Reducer performs the task of summarizing the result. There can be multiple reducers to parallelize the aggregations.

Some other technologies for handling big data

1. Parallel Computing – It involves the processing data on several machines simultaneously, each running its own OS , memory, computation speed and works on different parts of data. The output is communicated via message passing. Thus parallel computing helps in reducing the time for analysis of big data greatly.
2. Distributed File System – One or more central servers store files that can be accessed, with proper authorization rights, by any number of remote clients in the network. The distributed system uses a uniform naming convention and a mapping scheme to keep track of where files are located. When the client device retrieves a file from the server, the file appears as a normal file on the client machine, and the user is able to work with the file in the same ways as if it were stored locally on the workstation. When the user finishes working with the file, it is returned over the network to the server, which stores the now-altered file for retrieval at a later time.
3. Apache Hadoop - It is an open source software project that enables the distributed processing of large data sets across clusters of commodity servers. It can be scaled up from a single server to thousands of machines and with a very high degree of fault tolerance. Instead of relying on high-end hardware, the resiliency of these clusters comes from the software's ability to detect and handle failures at the application layer.
4. Data Intensive Computing – It is a class of parallel computing application which uses a data parallel approach to process big data. This works based on the principle of collocation of data and programs or algorithms used to perform computation. Parallel and distributed system of inter-connected stand alone computers that work together as a single integrated computing resource is used to process / analyze big data.

## VI. MAPREDUCE FOR SELF ORGANIZING WEBSITES : A THOUGHT

This section gives some thought on the feasibility of MapReduce on self-organizing websites. Self-organizing websites are websites that can adjust themselves according to customer needs. Here is an idea how can MapReduce be applied to such websites: The Mapper takes log entry files as input and filters out the IP address of sender and the page accessed. This intermediate result is saved. The Reducer reads the output from mapper and counts the number of instances for each IP address, page name pair. Another reducer takes his as input and produces the top n pairs of IP address, page name combination. This top n result can be used to display to each user the most frequent pages he has seen so far for a particular website.

This can also be used for advertisements on websites. Based on the customer footprints, it shall be able to determine which advertisement the user is/are interested.

## VII. CONCLUSION

In this article an overview of big data has been given. It also states some of the advantages of big data in our society, some technologies being used. This paper also states how BigData can be applied to self-organizing websites which can be extended to the field of advertising in companies.

Though this article does not cover the topic entirely but an attempt has been made to cover major aspects of big data

.